\documentclass[sigconf,anonymous=false]{acmart}

\usepackage{algorithmic}
\usepackage{graphicx}
\usepackage{textcomp}
\usepackage{xcolor}
\usepackage{makecell}

\usepackage{multirow}
\usepackage{siunitx}
\usepackage{pifont}% http://ctan.org/pkg/pifont
\usepackage{bm}
\usepackage{xcolor}  
\usepackage{soul}

\usepackage[ruled,%
            linesnumbered]{algorithm2e}
\SetKwInput{KwInput}{Input}
\AtBeginDocument{%
  }

\copyrightyear{2024}
\acmYear{2024}
\setcopyright{acmlicensed}\acmConference[ISWC '24]{Proceedings of the 2024
ACM International Symposium on Wearable Computers}{October 5--9,
2024}{Melbourne, VIC, Australia}
\acmBooktitle{Proceedings of the 2024 ACM International Symposium on
Wearable Computers (ISWC '24), October 5--9, 2024, Melbourne, VIC, Australia}
\acmDOI{10.1145/3675095.3676607}
\acmISBN{979-8-4007-1059-9/24/10}
\usepackage{fancyhdr}   %Kopfzeilen und co.
\fancypagestyle{firstpage}{     % define a custom header (Kopfzeile)
  \fancyhf{}
  \chead{\textcolor{gray}{This article has been accepted for publication in the Adjunct Proceedings of the 2024 ACM International Joint Conference on Pervasive and Ubiquitous Computing (UbiComp 2024) \& the 2024 ACM International Symposium on Wearable Computing (ISWC2024)}}
  \fancyfoot[C]{\small{\textcolor{gray}{~\copyright~ Personal use of this material is permitted.  Permission from ACM must be obtained for all other uses, in any current or future media, including reprinting/republishing this material for advertising or promotional purposes, creating new collective works, for resale or redistribution to servers \\ or lists, or reuse of any copyrighted component of this work in other works.}}}

}

\begin{document}

\title{On-device Learning of EEGNet-based Network For Wearable Motor Imagery Brain-Computer Interface }

\author{Sizhen Bian}
\orcid{0000-0001-6760-5539}
\affiliation{%
  \institution{PBL, D-ITET, ETH Zürich}
  \city{Zürich}
  \country{Switzerland}
}
\email{sizhen.bian@pbl.ee.ethz.ch}

\author{Pixi Kang}
\affiliation{%
  \institution{Department of Integrated Circuits, Tsinghua University}
  \city{Beijing}
  \country{China}}
\email{kpx21@mails.tsinghua.edu.cn}

\author{Julian Moosmann}
\affiliation{%
  \institution{PBL, D-ITET, ETH Zürich}
  \city{Zürich}
  \country{Switzerland}
}
\email{julian.moosmann@pbl.ee.ethz.ch}

\author{Mengxi Liu}
\affiliation{%
  \institution{German Research Center For Artificial Intelligence}
  \city{Kaiserslautern}
  \country{Germany}
}
\email{mengxi.liu@dfki.de}

\author{Pietro Bonazzi}
\affiliation{%
  \institution{PBL, D-ITET, ETH Zürich}
  \city{Zürich}
  \country{Switzerland}
}
\email{pietro.bonazzi@pbl.ee.ethz.ch}

\author{Roman Rosipal}
\affiliation{%
  \institution{Slovak Academy of Sciences}
  \city{Bratislava}
  \country{Slovak}
}
\email{roman.rosipal@savba.sk}

\author{Michele Magno}
\affiliation{%
  \institution{PBL, D-ITET, ETH Zürich}
  \city{Zürich}
  \country{Switzerland}
}
\email{michele.magno@pbl.ee.ethz.ch}

\renewcommand{\shortauthors}{Sizhen Bian et al.}

\begin{abstract}
Electroencephalogram (EEG)-based Brain-Computer Interfaces (BCIs) have garnered significant interest across various domains, including rehabilitation and robotics. Despite advancements in neural network-based EEG decoding, maintaining performance across diverse user populations remains challenging due to feature distribution drift. This paper presents an effective approach to address this challenge by implementing a lightweight and efficient on-device learning engine for wearable motor imagery recognition. The proposed approach, applied to the well-established EEGNet architecture, enables real-time and accurate adaptation to EEG signals from unregistered users. Leveraging the newly released low-power parallel RISC-V-based processor, GAP9 from Greeenwaves, and the Physionet EEG Motor Imagery dataset, we demonstrate a remarkable accuracy gain of up to 7.31\% with respect to the baseline with a memory footprint of 15.6 KByte. Furthermore, by optimizing the input stream, we achieve enhanced real-time performance without compromising inference accuracy. Our tailored approach exhibits inference time of 14.9 ms and 0.76 mJ per single inference and 20 us and 0.83 uJ per single update during online training. These findings highlight the feasibility of our method for edge EEG devices as well as other battery-powered wearable AI systems suffering from subject-dependant feature distribution drift.

\end{abstract}





\maketitle

\section{Introduction}

\thispagestyle{firstpage}

\begin{table*}[htbp]
\centering
\caption{On-device training methods on edge platforms}
\label{on_device}
\begin{tabular}{ p{0.8cm} p{3.6cm}  p{2.2cm} p{1.6cm} p{1.8cm} p{1.4cm}   p{3.4cm}  }
\hline
\textbf{Works} & \textbf{Learning Approaches} & \textbf{Usage scenario} & \textbf{Processing Device} & \textbf{Algorithm} & \textbf{Full Training}  & \textbf{Performance}\\
\hline
\cite{pellegrini2020latent} 2020&  CNN backprop. with  latent replay & image classification &  VEGA & MobileNetV1 \cite{howard2017mobilenets} &  no  & accuracy near the cumulative upper bound\\
\hline
\cite{ravaglia2021tinyml} 2021 & CNN backprop. with quantized latent replay & image classification &  Qualcomm Snapdragon & MobileNetV1 \cite{howard2017mobilenets}    & no  &  lossless compared with full-precision approach \\
\hline
\cite{ren2021tinyol} 2021 & Extended linear layer backprop. & anomaly detection/classification &  Arduino Nano & Autoencoder   & no  &  an acceptable
drop in accuracy of 10.7\% \\
\hline
\cite{avi2022incremental} 2022 & Expanded linear layer backprop. & hand gesture recognition with accelerometer &  STM32 & TinyOL \cite{ren2021tinyol} (v2)   & no  &  Online-training outperforms in F-score with scarce data pairs \\
\hline
\cite{craighero2023device} 2023 & Orchestrator-based forward/backward iteration & human activity recognition with inertial sensor &  STM32 & 1D-CNN   & yes  &  Full personalization of the CNN outperforms Transfer Learning in accuray\\
\hline
\cite{ficco2024federated} 2024 & Genann, Federated on-device training & ECG classification  &  ESP32, etc. & DeepEST   & yes  &  overall accuracy increased from 84.3\% to 86.22\%. \\
\hline
\cite{cioflan2024device} 2024 & Classifier layer(s) backprop. & keyword spotting  &  GAP9 & DS-CNN   & no  &  up to 14\% accuracy gain \\
\hline
\textbf{Ours} & Last dense layer backprop. & EEG motor imagery classification & GAP9 & EEGNet \cite{lawhern2018eegnet}    & no  &  7.31\% increase compared with baseline and real-time inference \\
\hline
\end{tabular}
\end{table*}

\begin{figure*}[hbt]
\graphicspath{{./Figures/}}
\centering
\includegraphics[width=0.80\linewidth, height = 3cm]{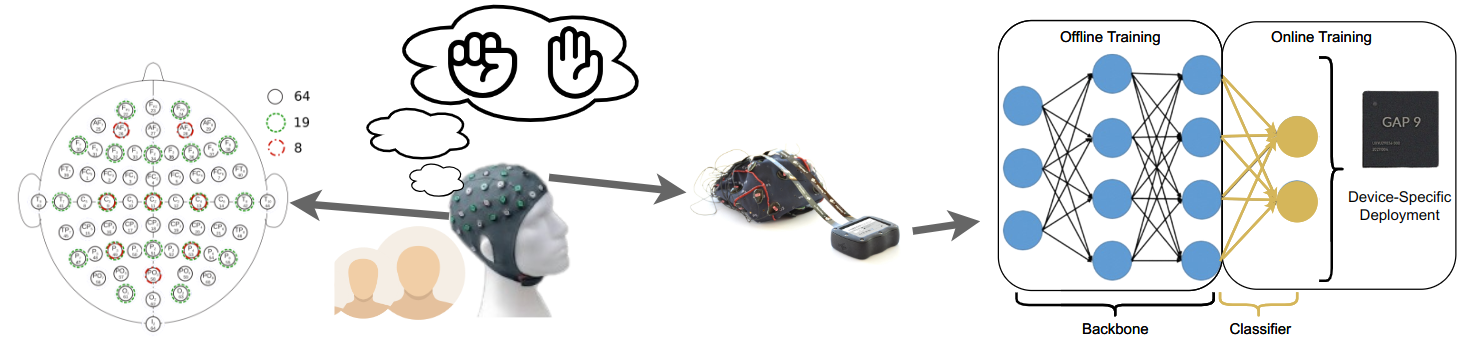}
\caption{Left: EEG electrode distribution and EEG-based motor imagery human-computer interface, which often faces the challenge of feature distribution drift across users and thus degrade the classification performance of a pre-trained neural network model;  Middle: wearable EEG device with an data processing unit at the extreme edge (image from g.tec GmbH); Right: the implemented dense layer (classifier) parameter updates through gradient descent-based backpropogation during online training, and targeted edge MCU: the GAP9 with RISC-V-based hardware accelerator}
\label{Overview}
\end{figure*}

Motor imagery brain-computer interface (MI-BCI) normally uses noninvasive electroencephalogram (EEG) electrodes to perceive a subject's intentions, aiming to present the kinetic movement in a digital system without the need to tense the muscles \cite{saibene2023eeg,liu2021tacnet}. Such an interface enables large-scale applications, e.g., for paralyzed users in rehabilitation \cite{alseydy2024using,anandika2023smart} and willingness control in robotics \cite{huang2021review}. For years, researchers have developed abundant algorithms to translate EEG signals into the subject intention, from traditional methods like kernel partial least squares (KPLS) decomposition \cite{trejo2015eeg}  or Riemannian covariances \cite{congedo2017riemannian}, to the emerged deep learning methods which abstract the EEG features via backpropagation enabled kernel optimization, such as convolutional neural network \cite{mao2020eeg} and transformer with self-attentions \cite{sun2021eeg}, etc., the later has proved to supply the state-of-the-art performance in motor imagery classifications  \cite{altaheri2023deep, hossain2023status}. However, one main challenge for EEG-based motor imagery classification, which still blocks the wide spread of consumer-level EEG systems, is the feature distribution drift of EEG signals across different subjects, caused by electrode/gel/skin interface \cite{kappenman2010effects}, physiological disturbance \cite{ribeiro2024breath}, or diverse personal nature \cite{matthews2017metrics, massullo2020abnormal}, etc., affecting the reusability and generalization of deep learning models. An intuitive method might be recollecting the training data and reconstructing the model, which is time and computationally costly, thus obstructing the model's wide-range deployment.

One main approach to address the feature distribution drift across users and avoid an extra expensive workload is the subject-specific network model based on transfer learning \cite{li2024transfer}, which transfers the knowledge learned in one group into the unregistered subjects. Such adaptive models maintain the interpreting ability of individuals with scarce data from the unregistered subjects. For example, in \cite{wang2020accurate}, the authors applied transfer learning on the Physionet EEG Motor Imagery dataset by first training and validating over the subjects via a five-fold cross-validation, then a subject-specific training and validation via a four-fold inter-subject generated the final customized models, achieving 70.8\% classification accuracy in a four-class task. In \cite{wei2021motor}, a deep transfer learning is developed and validated on dataset III of the second BCI competition, where the authors firstly apply the continuous wavelet transform (CWT) to convert the one-dimensional EEG signal into a two-dimensional time-frequency amplitude representation, and then a pre-trained convolutional neural network is further trained with scarce subject-specific EEG trials and utilized for two-class motor imagery classification, achieving a final classification accuracy of 96.43\%. Such works have demonstrated the effectiveness of subject-specific EEG motor imagery classification models based on transfer learning. However, a neglected problem is that those models need to be trained offline on a PC or cloud before the deployment. Such a train-then-deploy design process rigidly separates the learning phase from the runtime inference, resulting in complexity for end users \cite{bian2019wrist, bonazzi2023tinytracker}. In practical scenarios, especially on wearables, on-site adaptation is required to ease the post-training and deployment of the transferred models. In other words, the system should be capable of performing on-device continuous learning that automatically adapts to the ever-changing subjects or even the environment by only a few labeled input streams based on the original knowledge. Continuous learning aims to empower the system to incrementally learn non-stationary data streams, addressing the catastrophic forgetting issue that is commonly met in practical neural network deployments, such as testing samples from new tasks or data distribution \cite{adaimi2022lifelong, wang2024optimization}.

In this work, we experimentally show that by combining a novel processing unit, which incorporates parallel RISC-V Cores and an AI Hardware accelerator, and our proposed dedicated on-device learning engine, which is based on transfer learning, the accuracy of the EEG motion imaginary from unregistered users with scarce labeled input can be impressively improved, enabling on-site adaptation for wearable EEG systems.

\section{Related Work}

Addressing the performance degradation between registered and unregistered users has been an emerging research topic in the ubiquitous computing domain\cite{meegahapola2024m3bat, gong2024sotta}. While different strategies have been proposed \cite{wang2024optimization, niu2023towards}, a fundamental factor is neglected: the model adaptation, in a lot of cases, needs to be carried out locally on edge devices.
On-device learning enables models to train and update directly on extreme edge IoT devices continuously, which has already been successfully applied in different domains of interest. Table \ref{on_device} lists a few typical domain studies in recent years. The authors in \cite{pellegrini2020latent} introduced an on-device image classification model based on the MobileNetV1 \cite{howard2017mobilenets}. A method named "Latent Replay" was implemented by storing activation volumes at intermediate layers and achieved state-of-the-art performance on complex video benchmarks combined with continual learning techniques. An on-device training strategy for accelerometer-based hand gesture recognition was presented in \cite{avi2022incremental}, where a standard online training approach (TinyOL \cite{ren2021tinyol}) with minibatch-based backpropagation was implemented to deal with the catastrophic forgetting issue. When deploying the adaptive model on STM32, the authors observed an acceptable accuracy drop compared with the original models on unconstrained computing platforms. The same edge platform was also adopted in \cite{craighero2023device} for human activity recognition with a compact adaptive 1D-CNN model that could be trained at full scale. Three submodules were used to implement backpropagation via gradient descent, an orchestrator, and two forward and backward submodules. The orchestrator governs the iterative training procedure by invoking alternatively the Forward and the Backward sub-modules, and allows specifying training hyperparameters. The authors concluded that such a full personalization CNN even outperformed the utilized transfer learning method on the WISDM human activity recognition dataset.  In \cite{cioflan2024device}, a keyword spotting on-site adaptation system was developed to address accuracy degradation when neural networks are exposed to noisy environments. The system allows backpropagation-based weight optimization in the classifier layer(s) of a pre-trained neural network and achieves up to 14\% accuracy gains. On-device tests show that the adaptation on MCUs takes as little as 806 mJ in only 14 s. Besides above works, on-device training has also been explored by applying different schemes of backpropagation for anomaly detection \cite{ren2021tinyol}, ECG classification \cite{ficco2024federated} (with federated learning), regression application like fuel consumption prediction \cite{ficco2024federated}, etc. 

Up to date, there are few works exploring motor imagery with on-device training to overcome the feature drift of EEG signals among subjects, although such drift severely blocks the wide deployment of neural network-based EEG motor imagery classification models. In this work, we built a dedicated on-device training engine for a specific processing unit to enable online learning and demonstrated its efficiency by observing the classification performance with/without on-device training. In summary, we bring the following contributions in this work:

\begin{enumerate}
\item We propose an on-device learning engine based on dense layer backpropagation to adapt a subject-specific EEG motor imagery classification model. The model has been implemented and evaluated, showing a 7.31\% accuracy gain compared with online inference without subject-specific on-device training. 
\item To optimize the on-device training and inference performance regarding energy and time consumption, especially for wearable EEG devices, we scaled down the input stream (64 channels to 19 channels) and the network size (24.9 KByte to 15.6 KByte) without bringing a significant drop to the classification accuracy (56.98\% to 56.41\%). 
\item We presented a comprehensive evaluation of the computing efficiency of the on-device training and inference under different hardware hyper-parameters and input size configurations.

\end{enumerate}

\section{Methods}

Fig. \ref{Overview} illustrates a wearable EEG system that normally faces the issue of cross-subject feature distribution drift and the online training scheme we implemented in this work. The scheme is demonstrated on an IoT device targeting wearable systems for EEG motor imagery classification. In this section, we will describe all the materials and techniques adopted in this study.

\subsection{Dataset}

The well-known publicly available EEG Motor Imagery dataset, Physionet \cite{goldberger2000physiobank}, is used to verify our proposal since the dataset was collected from 109 volunteers (data from four of them are discarded as a result of variable trial number), which is suitable for online training evaluation. The volunteer performed different motor/imagery tasks while 64-channel EEG was recorded using the BCI2000 system with a sampling rate of 160Hz. Each volunteer performed 14 experimental runs: two one-minute baseline runs (one with eyes open, one with eyes closed) and three two-minute runs of each of the four following tasks:  open and close the left(L)/right(R) fist physically and mentally, open and close both fists/feet(F) physically and mentally. To keep in line with previous four-class classification studies (L/R/0/F where 0 means relax time) \cite{dose2018end, wang2020accurate}, we extracted the same classes from the raw dataset into instances of three seconds in length (480 samples) for exploration. Each run gives 21 trials per class per subject. Fig. \ref{Overview} (left) shows the spatial distribution of 64 electrodes attached across the scalp, which correspond to the 64 channels of the EEG signal. To evaluate the performance of input downscaling, which aims to access the edge training and inference in real-time, we also used 19 and 8 channels as the model input, as shown in the same figure.

%\begin{figure}[hbt]
%\graphicspath{{./Figures/}}
%\centering
%\includegraphics[width=0.65\linewidth, height = 4.0cm]{Figures/electrode.png}
%\caption{EEG Electrode}
%\label{electrode}
%\end{figure}

\subsection{Target device}

For wearable EEG-based system setup, we adopted the GAP9 microprocessor, a lowe power RISC-V-based PULP (Parallel Ultra Low Power) platform developed by GreenWaves Technologies, designed for ultra-efficient AI inference and signal processing. The GAP series is composed of three fundamental blocks: a fabric controller (FC), a smart peripheral controller (µDMA), and a parallel compute engine (Cluster). The FC is responsible for managing peripheral devices and controlling application execution on GAP9. The µDMA allows autonomous, low-energy peripheral management and on-the-fly calculation through specialized processing blocks for ultra-low latency tasks. The cluster provides a flexible, on-demand, high-performance programmable calculator for any task that demands significant compute resources such as digital signal processing or machine learning algorithms. Besides this, to achieve ultra-low power computing, the GAP series also adopts Dynamic Frequency and Voltage Scaling (DVFS) in multiple different domains of the chip, which allows elements of the chip to be entirely switched off when not in use but also the actual capabilities and energy consumption to be precisely tuned to the requirements of the task being executed. Such a heterogeneous design has been proven with impressive edge AI performance in a few case studies \cite{moosmann2023ultra, bonazzi2023low, moosmann2023flexible, bian2022exploring}.   

\subsection{Neural Network Model}

\begin{figure}[hbt]
\graphicspath{{./Figures/}}
\centering
\includegraphics[width=0.89\linewidth, height = 6.5cm]{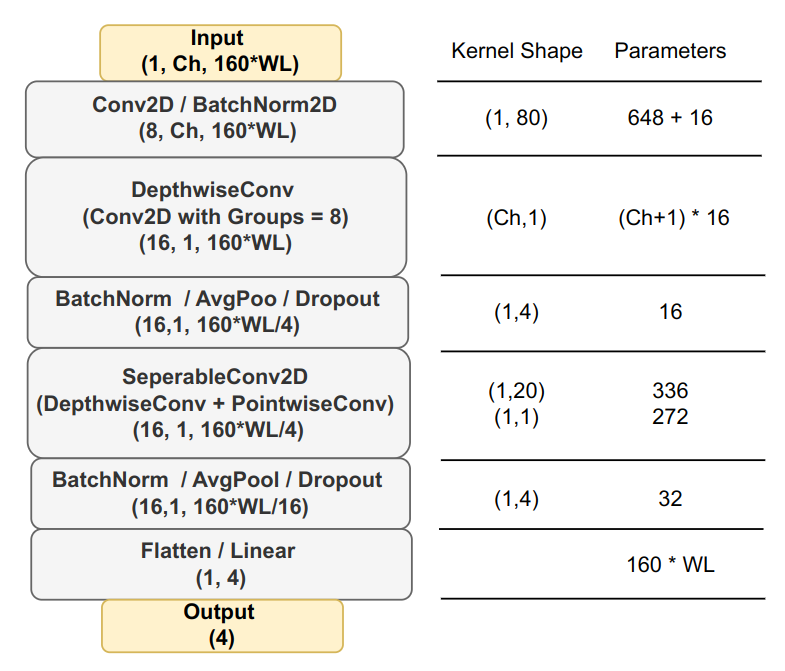}
\caption{Varied EEGNet applied in this work, where Ch means channel number (64, 19, 8), WL means window length (3s, 2s, 1s)}
\label{EEGNet}
\end{figure}

The motor imagery classification model applied in this work is the well-known EEGNet \cite{lawhern2018eegnet} (with slight adaption), a compact CNN architecture for EEG-based BCIs, proposed by Lawhern et al. in 2018. The network starts with a temporal convolution to abstract shallow features with eight kernels of size (1, SamplingRate/2), then a depthwise convolution is connected to each feature map individually with 16 kernels of size (ChannelNumber, 1), aiming to perceive the frequency-specific spatial patterns. The separable convolution (composed of a depthwise convolution with 16 kernels of size (1, SamplingRate/8) and a pointwise convolution) mixes the feature maps together after learning a temporal summary for each feature map individually. Although there are modified EEGNet for motor imagery recognition with higher accuracy\cite{rao2024optimized, lin2024eegnet}, we stick to EEGNet as first it is lightweight for edge systems and IoT devices, second, we are more interested in the performance gain instead of the absolute performance. The first variation of our model compared to the original EEGNet is that we used a fully connected layer followed by a softmax activation, acting as the classifier, gives the probability distribution of the four classes; while in the original model, the authors omit the use of a dense layer for feature aggregation prior to the softmax classification layer, aiming to reduce the number of free parameters in the model. The second variation is that we reduce the second average pooling size from (1, 8) to (1, 4), aiming to supply more space for on-device updating of the dense layer. As the original network was written in the Tensorflow framework, we first rebuilt the network with Pytorch to ease the following hardware deployment on the target device. The depthwise convolution is implemented by setting the group variable of the Conv2D function in Pytorch, as there is no direct depthwise convolution function in Pytorch. During training, the Adam scheme was used as the optimizer and initialized with a learning rate of 0.01 and decays with a gamma of 0.1 every 40 steps. An early stopping scheme is applied with a patience of 10. The categorical cross-entropy loss worked as the loss function in both offline and online training. For offline training, we set the batch size to 16, while during online training, it was set to 1. Fig. \ref{EEGNet} presents the building blocks of the network in Pytorch and the corresponding input/output sizes, kernel shape, and the trained amount of parameters. Besides the baseline with a channel number of 64 and a window length of 3 seconds, we also evaluated two model input sizes with channel numbers of 19 and 8 and window length of 2 and 1 second, which also scales down the network footprint from 24.9 KByte to 15.6 KByte and 8.5 KByte. For both training and testing online and offline, network parameters are kept as float type.

\subsection{Online training engine}

After the pre-trained model is deployed onto MCUs, the parameters of the backbone (the layers before the final linear one) are frozen and transferred for online adaptation with unregistered users, while the weights and bias parameters of the linear layer are updated iteratively using the gradients calculated on the device along the input stream and the corresponding label. The gradients needed for parameter update are obtained by taking derivatives of the log loss. To expedite the learning process and to escape from local minima, we implement Stochastic Gradient Descent (SGD) with momentum as the optimizer to manage past gradients and perform parameter updates during online training. Concretely, the exponential moving averages (EMA) of the parameters' historical gradients are maintained and updated each time there inputs a new sample. The proposed on-device algorithm \ref{algo} describes the online training pipeline and is implemented in C language for the target device.

\begin{algorithm}[]
  \SetAlgoLined
  \KwInput{\textrm{labeled EEG instances (x/y)}}
  \KwResult{\textrm{updated weight($w^u$) and bias($b^u$)}}
  Activate the online training engine by external command;
  \For{n \textbf{in} StreamingData}{
     $x^\mu_n$ = $Backbone(x_n)$\; 
     \textrm{Probability vector: }$p^\mu_n$ = $Classifier(x^\mu_n)$\; 
     \textrm{Cross-entropy loss: } $L^\mu_n$ = $LogLoss(onehot(y_n), p^\mu_n)$\; 
     \textrm{Gradient: }$[g^\mu_w, g^\mu_b]$ = $EMA(Deviation(L^\mu_n)$)\;
     \textrm{Update: } $[w^u, b^u]$ = $[w^u, b^u]$ - $\lambda[g^\mu_w, g^\mu_b]$ $(\lambda:$ \textrm{LearningRate}$)$\;
  }
  %return $[w^u, b^u]$ \;
  return $[w^u, b^u]$ \;

  \caption{On-device training pipeline}
  \label{algo}
\end{algorithm}

\begin{figure}[hbt]
\graphicspath{{./Figures/}}
\centering
\includegraphics[width=0.79\linewidth, height = 4.0cm]{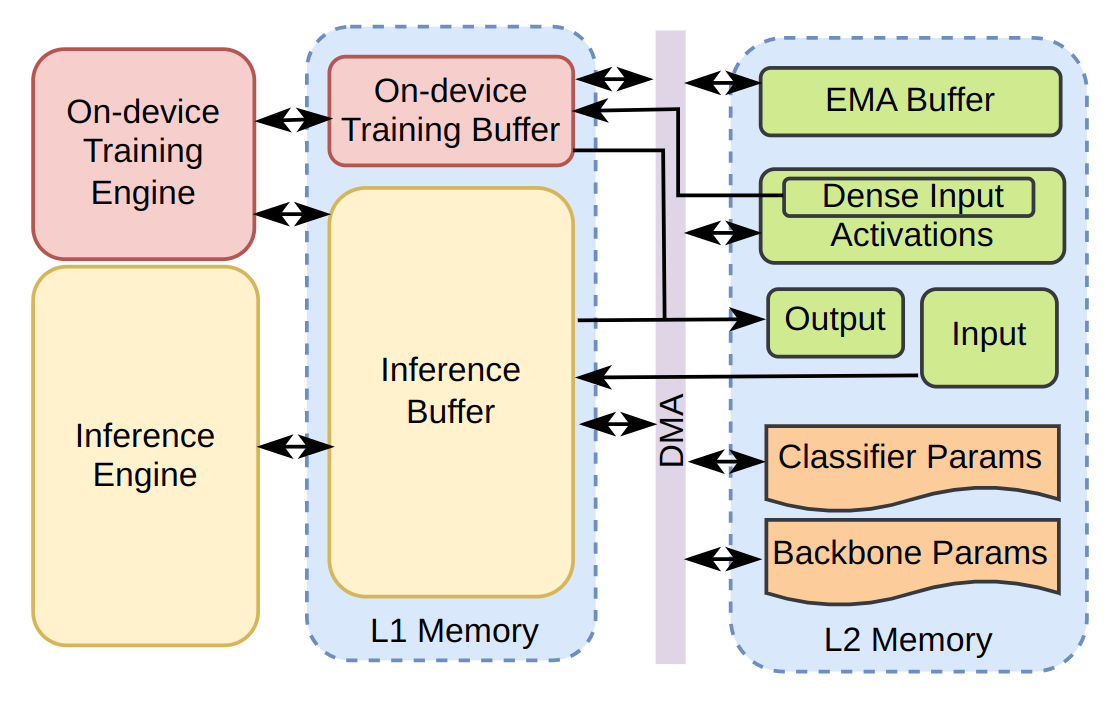}
\caption{Training Engine}
\label{Engine}
\end{figure}

Fig. \ref{Engine} depicts the on-board implementation structure for the task. The buffers for the backbone/classifier parameters, activations, input/output, and EMA are initialized and allocated in L2 memory, which contains a 1.6MiB SRAM memory. The network run is delegated to the cluster, where nine identical RISC-V cores embedded with FPU are deployed, sharing an L1 Tightly Coupled Data Memory (TCDM) of size 128KiB. For regular inference runs, data transfer patterns and required memory are automatically generated by the development framework, while for the on-device training, the data flow and memory structure involved are manually designated. During this time, the classifier parameters, input/output tensor of the classifier, and the EMA are transferred from L2 memory to L1 memory through DMA, updated in parallel by the cluster, and written back L2 memory after the training. In case the L1 memory can not satisfy the requirement, a tiled method is applied to balance the memory occupation and latency \cite{2023colibriuav}.

\section{Evaluation}

As described in the last section, we use EEGNet for the on-device training evaluation, which has been widely applied in related works with variations along improved capability and expanded volume to boost the motor imagery classification performance \cite{deng2021advanced, liu2022end, riyad2020incep}. Since the deployment in this work targets extreme edge IoT MCUs, we use the baseline of the EEGNet to avoid conflict with the memory budget. Meanwhile, to pursue edge computing efficiency regarding latency and energy consumption, we also evaluate the on-device training performance with another two model configurations by scaling down the input size and sampling window. Specifically, we chose the input channels of 19 and 8 and the sampling window of 2 and 1 second; the trainable parameters amount changes correspondingly. No quantization or pruning is used for compressing.

\subsection{Feature distribution drift evaluation}

Feature distribution drift often occurs when meeting users whose data are unavailable during offline training, resulting in a degradation of classification performance. 
To evaluate this degradation in the case of motor imagery and also the performance boosted by online training, we trained two networks using a small-scale dataset with instances from only 15 volunteers. The aim of such a split is to maximally simulate the practical scenarios and exploit the potential of online learning in cases where data collection is costly. The left volunteer's dataset was used for online training and online testing, with a split rate of 0.5. The two networks were trained with the following strategies:

\begin{enumerate}
\item Leave one user out: instances from three volunteers are selected as the validation dataset, while the instances from the other volunteers compose the training set, and five-fold cross-validation is used to train the model. 
\item Leave one session out: instances of each volunteer are shuffled and split into five sessions of equal-size sub-collections. Five-fold cross-validation is carried out: in each round, one session from each volunteer is selected as the validation set, while the union of the rest sessions is used as the training set.
\end{enumerate}

The aim of such a training strategy is to assess the influence of user-specific properties on the model performance, which can be expressed by the difference in the classification accuracy between the two trained models \cite{ferrari2020personalization, bian2022using}.
As Table \ref{Degradation} presents, the training accuracy of Leave-One-User-Out degrades with an accuracy of  5.43\% compared to the accuracy of Leave-One-Session-Out, which means that meeting the data from the unregistered users will weaken the motor imagery recognition ability to a certain extent, inferring the necessary of using online training to provide customized EEGNet for motor imagery classification. 

\begin{table}[hbt]
    \centering
    \caption{Classification drop evaluation}
    \label{Degradation}
    \begin{tabular}{p{1.5cm} p{1.5cm}  p{1.7cm}  p{1.5cm}  }
    \hline
    %\rule{0pt}{9pt}
     Performance    &  Leave-One-User-Out & Leave-One-Session-Out & Degradation \\
    \hline
    Accuracy     &  50.07\% &  55.5\%  & 5.43 \%  \\
    \hline
    \end{tabular}
%\begin{minipage}{\textwidth}
%    \setlength{\columnsep}{1.8cm}
%    $^{a}$ Leave-One-User-Out.
%\end{minipage}
\end{table}

\subsection{On-device training performance evaluation}

To evaluate the capability of post-deployment online training in model customization, half part of the instances from each remaining volunteer are used for online training, and another half part are used for online testing with and without online training. We use the averaged classification accuracy of all the remaining testers to assess the online testing performance before and after onine training. The Leave-One-User-Out offline trained model was used as the backbone, as it is more general in feature abstraction among trained users, and the other trained model is already prior to recognizing the motor imagery of the specific trained users. Table \ref{average_accuracy} listed the averaged inference accuracy with the three model configurations during different input statuses. 
As can be seen, the accuracy gain after the on-device training reaches up to 7.31\% when using 19 channels of EEG signal and with a window length of 2 seconds. With the even smaller input size (8 channels and 1 second window), there is still accuracy gain (5.87\%) by online training; however, the testing accuracy drops obviously, which is not the case when shrinking the input from 64 channels to 19 channels and window length from 3 to 2 seconds. Overall, the result indicates the efficiency of our proposal in EEGNet customization for boosting motor imagery recognition performance with new users.

\begin{table}[hbt]
    \centering
    \caption{Averaged inference accuracy with different model configurations}
    \label{average_accuracy}
    \begin{tabular}{p{1.0cm}  p{2.0cm} p{2.0cm} p{2.0cm} }
    \hline
    %\rule{0pt}{9pt}
    Model     &   online testing$^{a}$ & online testing$^{b}$ & accuracy gain$^{c}$  \\
    \hline
    Baseline     & 50.24\% & 56.98\% & \textbf{6.74\%}  \\
    \hline
    C\_One$^{d}$    &  49.10\% & 56.41\% & \textbf{7.31\%}  \\
    \hline
    C\_Two$^{e}$   &  46.67\% &  52.54\% & \textbf{5.87\%}\\
    \hline
    \end{tabular}
\begin{minipage}{\textwidth}
    \setlength{\columnsep}{1.8cm}
    $^{a}$ without online training.
    $^{b}$ with online training. \\
    $^{c}$ Gain of online testing after on-device training. \\
    $^{d}$ 19 channels, 2s window length (320 samples per instance). \\
    $^{e}$ 8 channels, 1s window length (160 samples per instance).
\end{minipage}
\end{table}

\subsection{On-device computing efficiency evaluation}

\begin{figure}[hbt]
\graphicspath{{./Figures/}}
\centering
\includegraphics[width=0.79\linewidth, height = 4.5cm]{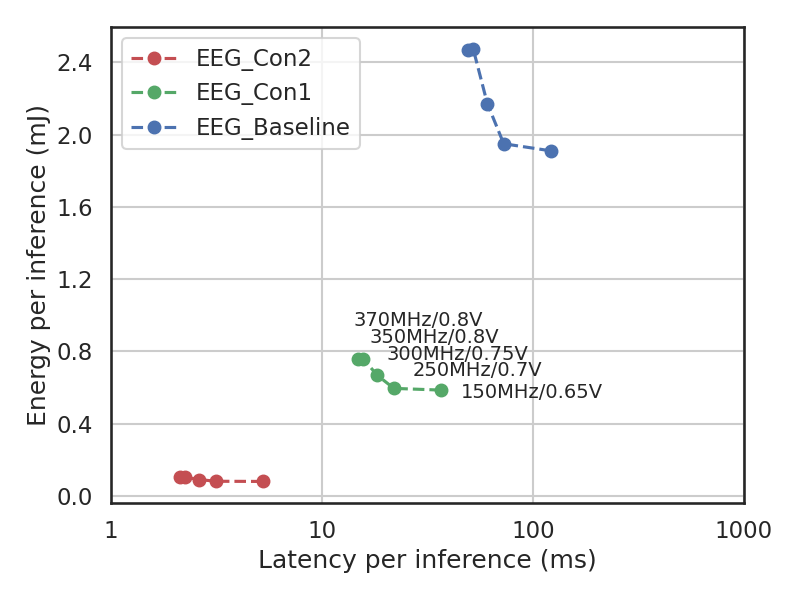}
\caption{Time and energy consumed per inference}
\label{infernceEff}
\end{figure}

\begin{figure}[hbt]
\graphicspath{{./Figures/}}
\centering
\includegraphics[width=0.79\linewidth, height = 4.5cm]{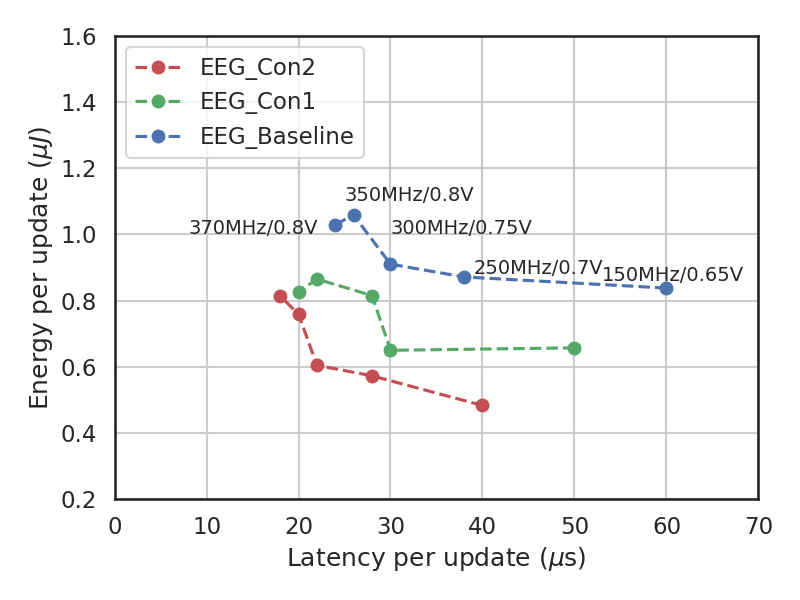}
\caption{Time and energy consumed per update}
\label{updateEff}
\end{figure}

The computing efficiency is evaluated by observing the on-device inference/training latency and energy consumption. We set five sets of clock speeds and input voltages to GAP9 for this evaluation: 150MHz/0.65V, 250MHz/0.7V, 300MHz/0.75V, 350MHz/0.8V, and 370MHz/0.8V, a power analyzer is used for the measurement. Fig. \ref{infernceEff} and Fig. \ref{updateEff} show the result. During the inference, a smaller input size costs less energy and inference time. In the case of 370 Mhz and 0.8 V power input, per inference takes 2.1 ms and 0.10 mJ, 14.9 ms and 0.76 mJ, 49.3 ms and 2.47 mJ for the three models with different input sizes, respectively. Considering the competitive online testing accuracy after online training, the model with a 2 s window size and 19 channels supplies a maximal 67 Hz update rate at the classification output. Since the online training only updates the dense layer parameters, the cost regarding latency and energy is much smaller than a regular inference. For example, with the 370 MHz and 0.8 V power input, the latency and energy consumption for the three models are 18 us and 0.81 uJ, 20 us and 0.83 uJ, 24 us and 1.03 uJ per update (with single sample and single epoch). Such results guarantee the feasibility of online training for battery-powered wearable EEG devices.

\section{Conclusion}

This work described an on-device training engine for the EEG-based motor imagery brain-computer interface targeting the IoT MCUs, aiming to address the recognition degradation caused by feature distribution drift in EEG signals when a pre-trained model meets unregistered users. We utilized the baseline of EEGNet, the well-known Physionet EEG dataset, and a newly released MCU with parallel cores for AI execution, demonstrated the effectiveness of our on-device training engine, where the parameters of the final linear layer of the EEGNet are updated along with inputting stream of unregistered users. The updating is supported by the standard backpropagation with stochastic gradient descent as the optimizer. We used three model configurations with different input sizes to assess the on-board inference and training performance, including the recognition ability and computing efficiency. The on-board evaluation shows that with on-device training, the customized EEGNet could bring a classification accuracy gain of up to 7.31 \% with 19 channels of EEG input and 2 s of window length. Regarding computing efficiency, the onboard experiment shows 14.9 ms and 0.76 mJ per inference and 20 us and 0.83 uJ per update with the same model configuration when giving 0.8 V power input and 370 MHz main clock speed to the device, indicating the feasibility of our online training proposal for edge EEG devices like the wearable form EEG devices powered by batteries.

\begin{acks}
This work was supported by the CHIST-ERA project ReHab(20CH21-203783).
\end{acks}
\bibliographystyle{ACM-Reference-Format}
\balance
\bibliography{sample-base}

%\appendix

\end{document}